\documentclass[prl,aps,twocolumn,amsfonts,showpacs,superscriptaddress,preprintnumbers]{revtex4-1}
\pdfoutput=1
\usepackage{graphicx}
\usepackage[utf8]{inputenc}
\usepackage{flushend}
\usepackage{dcolumn}
\usepackage{bm}
\usepackage{balance}
\usepackage[normalem]{ulem}
\usepackage[colorlinks = true,
            linkcolor = blue,
            urlcolor  = blue,
            citecolor = blue,
            anchorcolor = blue]{hyperref}
\usepackage{verbatim,subfigure}
\usepackage{color,ulem}
\usepackage[english]{babel}
\usepackage{MnSymbol,wasysym}
\usepackage[utf8]{inputenc}
\input Starburst.fd
\newcommand*\initfamily{\usefont{U}{Starburst}{xl}{n}}\initfamily 

\newcommand{\beq}{\begin{eqnarray}}
\newcommand{\eeq}{\end{eqnarray}}
\usepackage{amsmath}
\usepackage{tikz}
\usetikzlibrary{decorations.pathmorphing}
\usetikzlibrary{shapes.misc}
\tikzset{cross/.style={cross out, draw=black, minimum size=8*(#1-\pgflinewidth), inner sep=0pt, outer sep=0pt},
cross/.default={1pt}}
\usetikzlibrary{patterns,math}
\definecolor{applegreen}{rgb}{0.55, 0.71, 0.0}
\newcommand{\dd}{\mathrm{d}}

\begin{document}

\preprint{\texttt{IFT-UAM/CSIC-23-117}}

\title{Thermal Diffusivity and Pole-Skipping in the Incoherent Semilocally Critical IR}

\author{Hyun-Sik Jeong}

\email{hyunsik.jeong@csic.es}

\affiliation{Instituto de F\'isica Te\'orica UAM/CSIC, Calle Nicol\'as Cabrera 13-15, Cantoblanco, 28049 Madrid, Spain}
\affiliation{Departamento de F\'isica Te\'orica, Universidad Aut{\'o}noma de Madrid, 28049 Madrid, Spain}

\begin{abstract}
Pole-skipping offers compelling evidence for the hydrodynamic origin of chaotic behavior in strongly coupled quantum systems. We demonstrate that the cumulative effect of higher-order corrections to the hydrodynamic diffusive mode, captured by the parameter $\Omega$, determines the thermal diffusivity through chaos parameters, providing new insights into the interplay between chaos and hydrodynamics. In the incoherent limit, where momentum relaxation is pronounced, we show that the thermodynamically stable phase corresponds to a semi-locally critical IR fixed point. This finding extends previous analyses of simple IR fixed points and may offer a gravity dual description for stable phases beyond the Sachdev-Ye-Kitaev model in the incoherent regime. Additionally, we connect our findings to a universal bound on the thermal diffusion constant in holography, establishing a direct link with the dynamical critical exponent $z$. We derive a new relation, $\Omega = (2-z)/(2z-2)$, and propose that it characterizes the thermodynamically stable phase for generic values of $z$.
\end{abstract}

\maketitle

%%%%%%%%%%%%%%%%%%%%%
%
%%%%%%%%%%%%%%%%%%%%%
{\bf Introduction.---}
Chaos is a ubiquitous phenomenon that arises widely in nature. At the classical level, chaos is thought to underlie the microscopic dynamic processes governing macroscopic phenomena like transport and thermalization~\cite{Ott,Gaspard}. Nevertheless, whether chaos plays a fundamental role in quantum many-body systems remains an open question. There are promising indications that out-of-time ordered correlation functions (OTOCs)~\cite{Larkin,Maldacena:2015waa} -- the probe to study chaotic behavior with Lyapunov exponent $\lambda_L$ and the butterfly velocity $v_B$ -- is closely connected to transport and hydrodynamic behavior~\cite{Gu:2016oyy,Davison:2016ngz,Patel:2016wdy,Blake:2016wvh,Blake:2016sud,Blake:2016jnn,Blake:2017qgd,Grozdanov:2017ajz,Blake:2017ris,Grozdanov:2018atb,Lucas:2017ibu,Hartman:2017hhp,Davison:2018ofp}.

Given that hydrodynamic degrees of freedom are universal across various quantum many-body systems, it is tempting to explore a hydrodynamic origin for chaotic behavior, including OTOCs. Significant progress in this area has shown that quantum chaos can also appear in the thermal energy density two-point functions~\cite{Grozdanov:2017ajz,Blake:2017ris}, a phenomenon known as \textit{pole-skipping}, where it is established with an effective field theory for maximally chaotic systems~\cite{Blake:2017ris}.

Pole-skipping occurs at a specific momentum space point, $\left(\omega_{*}, k_{*}\right)$, where the energy density two-point function (EDTF) becomes ill-defined:
\begin{align}\label{SPP}
\begin{split}
\omega_{*} = i \lambda_L \,,\quad k_{*} = i \frac{\lambda_L}{v_B} \,.
\end{split}    
\end{align}
At this point, the residue of the two-point function vanishes, ``skipping" the pole. Pole-skipping has been confirmed through near-horizon analyses of Einstein's equations and observed in various systems, including  Sachdev-Ye-Kitaev (SYK) models~\cite{Gu:2016oyy}, large central charge two-dimensional conformal field theories~\cite{Haehl:2018izb}, and across multiple contexts~\cite{Blake:2018leo,Grozdanov:2019uhi,Blake:2019otz,Natsuume:2019xcy,Natsuume:2019sfp,Natsuume:2019vcv,Ceplak:2019ymw,Ahn:2019rnq,Ahn:2020bks,Ramirez:2020qer,Ahn:2020baf,Natsuume:2020snz,Ceplak:2021efc,Jeong:2021zhz,Natsuume:2021fhn,Blake:2021hjj,Jeong:2022luo,Wang:2022mcq,Amano:2022mlu,Yuan:2023tft,Grozdanov:2023txs,Natsuume:2023lzy,Ning:2023ggs,Grozdanov:2023tag,Jeong:2023rck,Natsuume:2023hsz}.

Pole-skipping not only offers a universal method for computing $\lambda_L$ and $v_B$ from the EDTF, bypassing the need for OTOCs calculations~\cite{Shenker:2013pqa,Roberts:2014isa,Roberts:2016wdl,Shenker:2013yza,Shenker:2014cwa}, but also provides evidence for the hydrodynamic origin of chaotic behavior~\cite{Grozdanov:2017ajz,Blake:2017ris,Blake:2018leo}. This is seen in the fact that the poles of EDTF passes through the point \eqref{SPP} as
\begin{align}\label{SPPseries}
\begin{split}
\omega_{*} \equiv \omega (k_{*}) = -i\, D_T\, k_{*}^2 - i\, \sum\limits_{n=2}^{\infty}\, c_{2n}\, k_{*}^{2n} \,,
\end{split}    
\end{align}
where the diffusion mode is an infinite Puiseux series in $k^2$~\cite{Grozdanov:2019kge,Grozdanov:2019uhi}, with $D_T$ as the thermal (energy) diffusion constant and $c_n$ representing all higher-order coefficients.

We shed the new light on the intricate relation between transport, hydrodynamics, and chaos by a significant consequence of \eqref{SPPseries}:  
\begin{align}\label{SPPseries2}
\begin{split}
D_T \, \frac{\lambda_L}{v_B^2} = 1 + \sum\limits_{n=2}^{\infty}\, c_{2n}\,\frac{(-1)^{n} \, \lambda_L^{2n-1}}{v_B^{2n}}  \equiv 1 + \Omega \,,
\end{split}
\end{align}
where the cumulative effect of higher-order corrections, $\Omega$, reveals how intricate interactions influence the system's diffusion properties. This is particularly important in strongly coupled systems, critical phenomena, and quantum critical points, where such effects may significantly impact the system's behavior.

In the literature~\cite{Blake:2017ris,Blake:2018leo,Jeong:2021zhz,Jeong:2022luo}, it has been shown that particle-hole symmetric theories yield $\Omega=0$ in the incoherent limit, where the infrared (IR) fixed point is AdS$_2\times \mathbb{R}^2$, resulting in $D_T \, {\lambda_L}/{v_B^2} = 1$ which matches the finding in SYK models~\cite{Davison:2016ngz,Gu:2016oyy,Gu:2017njx}.

In this Letter, we demonstrate through an examination of thermodynamic instability that this is not the case; the AdS$_2 \times \mathbb{R}^2$ fixed point is actually unstable in the incoherent regime. Instead, the stable phase corresponds to another IR fixed point, known as the conformal to AdS$_2 \times \mathbb{R}^2$, with a finite value of $\Omega = -1/2$.

In the incoherent regime, characterized by strong momentum relaxation, it has been argued~\cite{Hartnoll:2014lpa} that extrinsic effects are eliminated, allowing transport to be governed by intrinsic and universal properties, as evidenced by phenomena such as linear-$T$ resistivity in holographic strange metals~\cite{Davison:2013txa,Jeong:2018tua,Ahn:2023ciq} and a universal bound on the thermal diffusion constant~\cite{Gouteraux:2014hca,Blake:2016wvh,Blake:2016sud,Blake:2017qgd,Jeong:2017rxg,Kim:2017dgz}.

Our analysis provides an explicit proof of concept that higher-order terms are not negligible but are crucial for accurately describing the thermodynamically stable system, particularly in extreme regimes exhibiting universal features of strongly coupled systems. It also suggests the need for alternative models beyond SYK in the incoherent regime.

%%%%%%%%%%%%%%%%%%%%%
%
%%%%%%%%%%%%%%%%%%%%%
{\bf The holographic Gubser-Rocha model.---}
We consider a holographic Einstein-Maxwell-dilaton-axion model known as the Gubser-Rocha model~\cite{Gubser:2009qt}, 
\begin{equation}\label{GRMA}
\begin{split}
S = \int \dd^4x\sqrt{-g}\, \bigg[ & R-\frac{1}{4} e^{\frac{\phi}{\sqrt{3}}}  F^2 -\frac{1}{2}(\partial{\phi})^2 + 6\cosh \frac{\phi}{\sqrt{3}} \\
& - \frac{1}{2} \sum_{I=x,y}(\partial \psi_{I})^2  \bigg] \,,
\end{split}
\end{equation}
where it includes a $U(1)$ gauge field with field strength $F=\dd A$ and a dilaton field $\phi$. Axion fields $\psi_I$~\cite{Andrade:2013gsa} is also added to introduce momentum relaxation in the boundary theory.

This model is an effective holographic model for condensed matter systems~\cite{Charmousis:2010zz,Gouteraux:2014hca}. 
Its IR scale invariant point falls into the class of the semi-local quantum liquids~\cite{Iqbal:2011in} and captures the characteristics of strange metals \footnote{For instance, it describes linear-$T$ resistivity, fermionic spectral function, and Homes's law in high-$T_c$ superconductors~\cite{Davison:2013txa,Anantua:2012nj,Jeong:2018tua,Ahn:2019lrh,Jeong:2019zab,Jeong:2021wiu,Balm:2022bju}. See also \cite{Ahn:2023ciq} for its limitation to describe the transport anomalies such as the Hall angle.}. 

We will show that the Gubser-Rocha model with particle-hole symmetry, vanishing chemical potential, exhibits two competing phases: the conformal to AdS$_2$ phase and the AdS$_2$ phase considered in the literature~\cite{Blake:2017ris,Blake:2018leo,Jeong:2021zhz,Jeong:2022luo}.

An intriguing aspect of the Gubser-Rocha model is its allowance for analytic background solutions. For the metric,
\begin{equation}\label{}
\begin{split}
\dd s^2 &=  -f(r) \dd t^2 + \frac{1}{f(r)} \dd r^2 + h(r) \left(\dd x^2 + \dd y^2\right) \,,
\end{split}
\end{equation}
where
\begin{equation}\label{ANSOL1}
\begin{split}
 f(r) &= r^2 \left[ 1 - \frac{m^2}{2(Q+r)^2} - \frac{(Q+1)^3}{(Q+r)^3} \left( 1- \frac{m^2}{2(Q+1)^2} \right)  \right]  \\
        & \quad \times \left(1+\frac{Q}{r}\right)^{\frac{3}{2}} \,, \qquad
 h(r) = r^2 \left(1+\frac{Q}{r}\right)^{\frac{3}{2}} \,,
\end{split}
\end{equation}
and the matter fields are 
\begin{equation}\label{ANSOL2}
\begin{split}
A &= \sqrt{3Q(Q+1)\left(1-\frac{m^2}{2(Q+1)^2}\right)} \left(1-\frac{Q+1}{Q+r}\right) \, \dd {t} \,, \\
\phi &= \frac{\sqrt{3}}{2} \log \left(1+\frac{Q}{r}\right) \,,\quad 
\psi_{x} = m \, {x} \,, \quad\,\, \psi_{y}=m \, {y} \,.
\end{split}
\end{equation}
Here, $r=1$ is the location of black brane horizon,  $m$ denotes the strength of momentum relaxation, and $Q$ is a parameter associated with the dilaton field.

%%%%%%%%%%%%%%%%%%%%%
%
%%%%%%%%%%%%%%%%%%%%%
{\bf Thermodynamic instability with particle-hole symmetry.---}
Particle-hole symmetric solutions are obtained from \eqref{ANSOL2}
\begin{equation}\label{LAvsGR}
\begin{split}
Q=0  \,, \qquad Q = \frac{m}{\sqrt{2}}-1 \,,
\end{split}
\end{equation}
where another solution $Q=-1$ is not allowed \footnote{See the details in \cite{Blake:2016jnn,Blake:2017qgd,Kim:2017dgz} also including the theories with broken particle-hole symmetry where the thermal diffusion sector is coupled with the charge diffusion sector.}.
 
The solution $Q=0$ corresponds to a vanishing dilaton field $\phi$, featuring an AdS$_2$ fixed point as studied in~\cite{Blake:2017ris,Blake:2018leo,Jeong:2021zhz,Jeong:2022luo}. The second solution involves a non-trivial dilaton field profile with a conformal to AdS$_2$ fixed point~\cite{Gubser:2009qt,Davison:2013txa}.

To determine which solution represents the ground state, 
we compute the difference in grand potential density $G$~\cite{Caldarelli:2016nni}:
\begin{equation}\label{GPDD}
\begin{split}
\Delta G \,\equiv\, G\,|_{Q=0} \,-\, G\,|_{Q=\frac{m}{\sqrt{2}}-1} \,,
\end{split}
\end{equation}
where
\begin{equation}\label{TPDGR}
\begin{split}
G = -\left(Q+1\right)^3 - \frac{m^2}{2} \,.
\end{split}
\end{equation}
Here, the thermodynamic potential density is considered at fixed chemical potential $\mu$ and temperature $T$~\cite{Caldarelli:2016nni}, expressed as $G=\epsilon-sT-\mu \rho$, where $\epsilon$ denotes the energy density, $s$ the entropy density, and $\rho$ the charge density. For the Gubser-Rocha model \eqref{GRMA}, $G$ corresponds to Eq. \eqref{TPDGR}: further details can be found in \cite{Kim:2017dgz}.

{
In the Gubser-Rocha model, the scalar field acquires a vacuum expectation value (VEV) while the corresponding source remains zero. This follows from the boundary conditions and the structure of the Fefferman-Graham expansion. Specifically, the scalar field near the AdS boundary takes the form $\phi = \phi_0 z + \phi_1 z^2 + \mathcal{O}(z^3)$, where the coefficients satisfy the condition ensuring the vanishing of the single-trace source. This establishes that the scalar operator can develop a VEV spontaneously. A detailed discussion of this result, including its derivation from the renormalized on-shell action and holographic Ward identities, can be found in Ref. \cite{Caldarelli:2016nni}.
}

\begin{figure}
\centering
    \includegraphics[width=\linewidth]{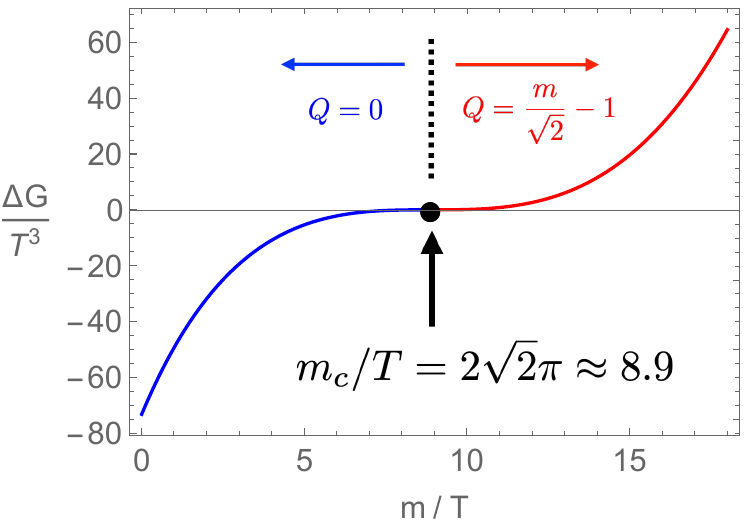}
\caption{The difference in grand potential density \eqref{GPDD}. $Q=0$ is the ground state for $m<m_c$ and $Q = \frac{m}{\sqrt{2}}-1$ is the one for $m>m_c$.}\label{GPDFIG}
\end{figure}

Within particle-hole symmetric theories, the only dimensionless parameter can be constructed using temperature $T=\frac{6(Q+1)^2-m^2}{8\pi (Q+1)^{3/2}}$. Then, the critical transition point ($m_c/T$) can be found analytically as
\begin{equation}\label{PTP}
\begin{split}
{m_c}/{T} = 2\sqrt{2} \pi \,,
\end{split}
\end{equation}
which occurs when $\Delta G=0$.
In figure \ref{GPDFIG}, we show $\Delta G/T^3$ vs. $m/T$, revealing that $Q=0$ is the ground state for $m<m_c$, while $Q = \frac{m}{\sqrt{2}}-1$ is the one for $m>m_c$.
Notably, the critical point corresponds to the phase at $m=\sqrt{2}$, which has been reported as an $\text{SL}(2,\mathrm{R}) \times \text{SL}(2,\mathrm{R})$ invariant point~\cite{Davison:2014lua}.

Our analysis concludes that when the momentum relaxation is strong ($m/T\gg1$) -- incoherent regime -- the conformal to AdS$_2$ fixed point is the thermodynamically stable solution over the AdS$_2$ fixed point discussed in \cite{Blake:2017ris,Blake:2018leo,Jeong:2021zhz,Jeong:2022luo}.

%%%%%%%%%%%%%%%%%%%%%
%
%%%%%%%%%%%%%%%%%%%%%
{\bf Quasi-normal spectrum and pole-skipping.---}
We analyze the higher-order corrections $\Omega$ in \eqref{SPPseries2} for the incoherent regime of Gubser-Rocha model. This involves studying the poles of EDTF \eqref{SPPseries}. If the pole-skipping point cannot be described by the relation $\omega_{*} = -i\, D_T\, k_{*}^2$, it indicates that $\Omega\neq0$.

To compute the quasi-normal spectrum of the EDTF, we consider bulk perturbations $\delta \mathcal{F}$ in the Fourier form 
$\delta \mathcal{F} = \delta \bar{\mathcal{F}}(r) \, e^{-i \omega t \,+\, i k x}.$
Using the standard holographic technique~\cite{Kovtun:2005ev} and taking advantage of diffeomorphism and gauge invariance, we derive the gauge-invariant variables $Z_i$ for the EDTF:
$Z_1 \equiv {4 k}/{\omega} \, \delta \bar{g}_t^x + 2 \delta \bar{g}_x^x - \left( 2 - {2 k^2 f'}/{(\omega^2 h')} \right) \delta \bar{g}_y^y + {2k^2 f}/{(\omega^2 h)} \, \delta \bar{g}_t^t ,\,
Z_2 \equiv \delta \bar{\psi}_x + {i m}/{(2 k)} \left( \delta \bar{g}_x^x - \delta \bar{g}_y^y \right) ,\,
Z_3 \equiv \delta \bar{\phi} - {h \phi'}/{h'} \, \delta \bar{g}_y^y \,.$
By solving the fluctuation equations of motion given in terms of $Z_i$, we construct the source matrix $\mathcal{S}$, composed of the leading coefficients of $Z_i$ on the boundary, i.e., Dirichlet boundary condition. The quasi-normal spectrum is then determined by finding the values of $(\omega, k)$ for which the determinant of $\mathcal{S}$ vanishes~\cite{Kaminski:2009dh}.

\begin{figure}
\centering
    \includegraphics[width=\linewidth]{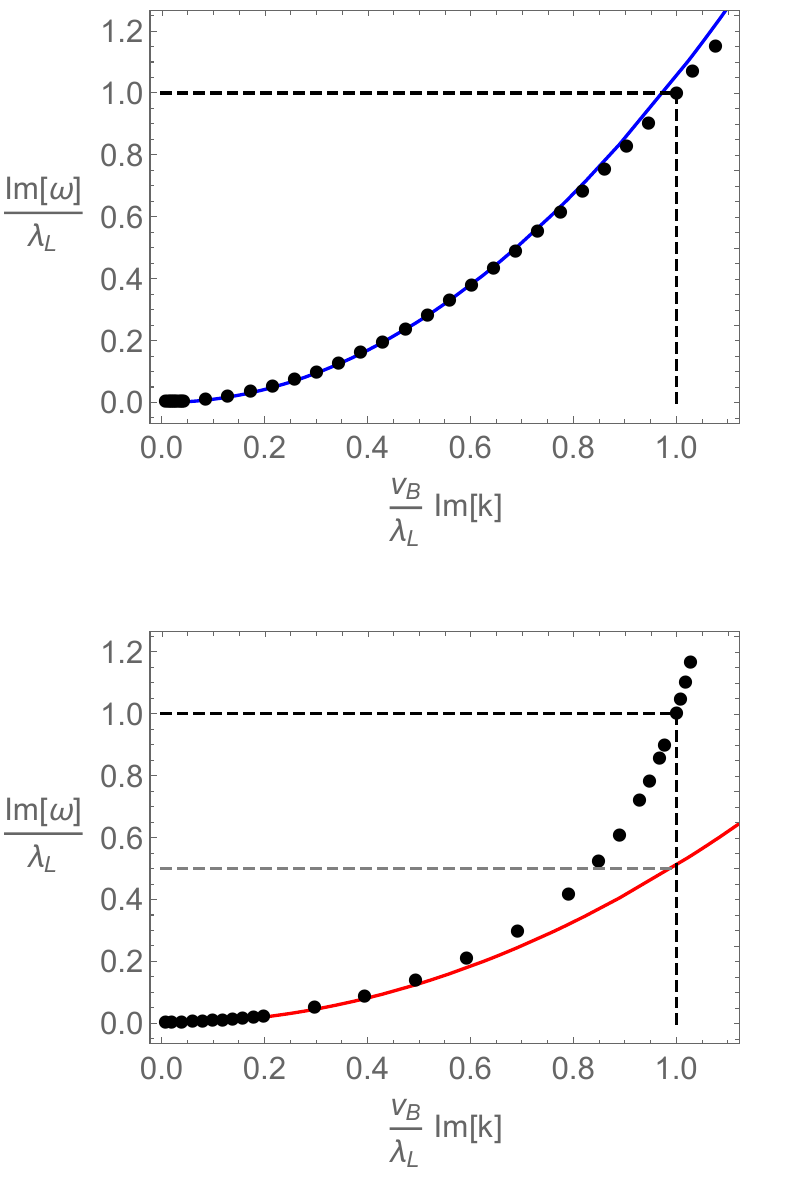}
\caption{Quasi-normal spectrum at $m/T=100$. The upper panel shows the case for $Q=0$, while the lower panel is for $Q = \frac{m}{\sqrt{2}}-1$. Dots are numerically obtained quasi-normal mode. Black dashed lines indicate the pole-skipping point \eqref{SPP} and solid lines are the diffusive mode $\omega(k) = - i D_T k^2$. Gray dashed line in the lower panel is \eqref{PSDIF22}.}\label{m100FIG}
\end{figure}

Figure \ref{m100FIG} shows the quasi-normal spectrum in the incoherent regime with $m/T=100$.
To interpret these results, we scale them with the chaos parameters ($\lambda_L$ and $v_B$)
\begin{equation}\label{}
\begin{split}
\lambda_L = 2\pi T \,, \quad v_B = \sqrt{\frac{2\pi T}{h'(1)}} \,,
\end{split}
\end{equation}
derived from the near-horizon analysis of pole-skipping in generic holographic Q-lattice models~\cite{Jeong:2021zhz}, consistent with the shock-wave analysis of OTOCs~\cite{Blake:2016wvh}.

For the case with $Q=0$ (upper panel), the higher-order corrections are negligible near the pole-skipping point, as reported in \cite{Blake:2017ris,Blake:2018leo,Jeong:2021zhz,Jeong:2022luo}.
Here, the lowest diffusive mode, $\omega_{*} = - i D_T k_{*}^2$, serves as a good approximation. The thermal diffusivity is defined via the Einstein relation $D_T = {\kappa}/{c_{\rho}}$, where $\kappa$ is the thermal conductivity, and $c_\rho = T (\partial s / \partial T)_\rho$ is the specific feat at fixed density $\rho$. For the Gubser-Rocha model with particle-hole symmetry, it can be computed as
\begin{align}\label{}
\begin{split}
D_T = 
\begin{cases}
\,\, \frac{\sqrt{6 m^2 + 16\pi^2 T^2}}{2m^2} \qquad\,\, (Q=0) \,, \\
\,\, \frac{4\pi T}{m^2} \qquad\qquad\qquad \left(Q=\frac{m}{\sqrt{2}}-1\right) \,.
\end{cases}
\end{split}
\end{align}

However, for $Q= \frac{m}{\sqrt{2}}-1$ (lower panel), the lowest diffusive mode does not intersect the pole-skipping point, indicating that higher-order corrections are necessary to accurately describe the pole-skipping point for the thermodynamically stable phase in the incoherent regime: we find a numerical fitting suggesting that corrections up to $n=6$ in \eqref{SPPseries} are required.

To extract the exact value of $\Omega$ in \eqref{SPPseries2}, we note that for $Q= \frac{m}{\sqrt{2}}-1$, the pole-skipping point is related to the ``modified" diffusive mode as
\begin{align}\label{PSDIF22}
\begin{split}
{\omega_{*}}/{2} = - i D_T k_{*}^2  \quad\longrightarrow\quad  {D_{T}\lambda_{L}}/{v_B^2} = {1}/{2} \,,
\end{split}    
\end{align}
as depicted by the gray dashed line in the lower panel, which can also be confirmed analytically as ${D_{T}\lambda_{L}}/{v_B^2} = {1}/{2} + 12 \pi^2 (T/m)^2 + \mathcal{O}(T/m)^4$ in the incoherent limit. Conclusively, this results in $\Omega = -1/2$ via \eqref{SPPseries2}, which is consistent with the numerical evaluation of our data at $m/T=100$: $\Omega = -0.500978$ with the series \eqref{SPPseries} up to $n=6$.

%%%%%%%%%%%%%%%%%%%%%
%
%%%%%%%%%%%%%%%%%%%%%
{\bf Discussion.---}
In this Letter, we have shown that pole-skipping provides concrete evidence for the hydrodynamic origin of chaotic behavior, highlighting that the cumulative effect of higher-order coefficients to the hydrodynamic mode, can determine thermal diffusivity through the chaos parameters, as seen in \eqref{SPPseries2}.

In particular, within the incoherent limit, where momentum relaxation is so strong that the system's intrinsic and universal properties dominate, we have proved that the thermodynamically stable phase corresponds to a semi-locally critical IR fixed point that is conformal to AdS$_2 \times \mathbb{R}^2$. Our findings imply that prior analyses of pole-skipping with a simple IR fixed point~\cite{Blake:2017ris,Blake:2018leo,Jeong:2021zhz,Jeong:2022luo}, such as AdS$_2 \times \mathbb{R}^2$, transition to a phase conformal to AdS$_2 \times \mathbb{R}^2$. This result may indicate a gravity dual description for the stable phase in the incoherent regime beyond the SYK model~\cite{Davison:2016ngz,Gu:2016oyy,Gu:2017njx}. It would be interesting to identify a corresponding ``extended" SYK-like model that exhibits this feature.

Additionally, we integrate the pole-skipping results from \eqref{SPPseries2} with \textit{the} celebrated universal bound on thermal diffusion constant~\cite{Gouteraux:2014hca,Blake:2016wvh,Blake:2016sud,Blake:2017qgd} in holography
$D_T \, {\lambda_L}/{v_B^2} = {z}/{(2z-2)} \,,$
which applies in the incoherent limit \footnote{The bound is not well defined at $z=1$ which presents an issue as it corresponds to an  irrelevant deformation of the quantum critical state. An independent and distinct analysis is required in this scenario as in \cite{Davison:2018nxm}.}. Here, a parametric cancellation between $D_T$ and ${\lambda_L}/{v_B^2}$ yields a quantity that depends solely on the IR parameter, the dynamical critical exponent $z$. Combining these expressions, we derive 
\begin{align}\label{SPPseries3}
\begin{split}
\Omega \,=\, \frac{2-z}{2z-2} \,.
\end{split}
\end{align}

Notably, the competing phases described in \eqref{LAvsGR} --one conformal to AdS$_2$ and the AdS$_2$ phase-- share the same IR exponent $z$ as $z \to \infty$ in the incoherent limit \footnote{In condensed matter systems, thermal diffusion bounds different from those found in holography are possible, such as in the models discussed in \cite{Patel:2016wdy}, which have $z = 3/2$ but show $D_T \, {\lambda_L}/{v_B^2} = 0.42$.}. Among these, only the thermodynamically stable phase, which is conformal to AdS$_2$, satisfies \eqref{SPPseries3}, where $\Omega = -1/2$ \footnote{The universal bound on thermal diffusion, $D_T \, {\lambda_L}/{v_B^2} = {z}/{(2z-2)}$, involves a diverging (running) dilaton at low temperatures in its derivation~\cite{Blake:2017qgd}, whereas the AdS$_2$ fixed point has a constant dilaton.}. Based on our findings, it is tempting to suggest that systems described by \eqref{SPPseries3} represent the thermodynamically stable phase even for other values of the IR fixed point parameter $z$. Investigating this possibility further would be an intriguing direction for future research.

Our findings provide a framework for understanding the behavior of strongly coupled quantum systems in the incoherent regime and suggest potential avenues for future exploration in related condensed matter models. One interesting test of our proposal for strongly coupled quantum systems may be the case of $z=2$, which leads to a vanishing $\Omega$. The value $z=2$ has previously appeared in high-$T_c$ superconductors~\cite{Abanov2003,Metlitski:2010aa,Patel:2014aa}.\\

%%%%%%%%%%%%%%%%%%%%%
%
%%%%%%%%%%%%%%%%%%%%%
{\bf Acknowledgements.---}
We would like to thank {Daniel Are\'an, Yongjun Ahn, Kyoung-Bum Huh, Viktor Jahnke, Juan F. Pedraza} for valuable discussions and correspondence. 
H.-S Jeong acknowledges the support of the Spanish MINECO ``Centro de Excelencia Severo Ochoa'' Programme under grant SEV-2012-0249. This work is supported through the grants CEX2020-001007-S and PID2021-123017NB-I00, funded by MCIN/AEI/10.13039/501100011033 and by ERDF A way of making Europe.

%%%%%%%%%%%%%%%%%%%%%
%
%%%%%%%%%%%%%%%%%%%%%
\bibliographystyle{apsrev4-1}
\bibliography{Refs}

\end{document}